\documentclass[preprint,showpacs,preprintnumbers,amsmath,amssymb]{revtex4}
\usepackage{booktabs}
\usepackage{mathrsfs}
\usepackage{epsfig}
\usepackage{graphicx}% Include figure files
\usepackage{dcolumn}% Align table columns on decimal point
\usepackage{bm}% bold math
\usepackage{amsmath}
\usepackage{slashed}       % slashed p

\let\jnfont=\rm
\def\NPB#1,{{\jnfont Nucl.\ Phys.\ B }{\bf #1},}
\def\PLB#1,{{\jnfont Phys.\ Lett.\ B }{\bf #1},}
\def\EPJC#1,{{\jnfont Eur.\ Phys.\ Jour.\ C }{\bf #1},}
\def\PRD#1,{{\jnfont Phys.\ Rev.\ D }{\bf #1},}
\def\PRL#1,{{\jnfont Phys.\ Rev.\ Lett.\ }{\bf #1},}
\def\MPLA#1,{{\jnfont Mod.\ Phys.\ Lett.\ A }{\bf #1},}
\def\JPG#1,{{\jnfont J.\ Phys.\ G}{\bf #1},}
\def\CTP#1,{{\jnfont Commun.\ Theor.\ Phys.\ }{\bf #1},}
\def\ZPC#1,{{\jnfont Z.\ Phys.\ C }{\bf #1},}
\def\JHEP#1,{{\jnfont JHEP \ }{\bf #1},}
\def\lsim{\raise0.3ex\hbox{$<$\kern-0.75em\raise-1.1ex\hbox{$\sim$}}}
\def\gsim{\raise0.3ex\hbox{$>$\kern-0.75em\raise-1.1ex\hbox{$\sim$}}}

\begin{document}

\title{Pair production of a 125 GeV Higgs boson in MSSM and NMSSM at the ILC}

\author{Zhaoxia Heng, Liangliang Shang, Peihua Wan }
\affiliation{
        Department of Physics,
        Henan Normal University, Xinxiang 453007, China
\vspace{1cm}}

\begin{abstract}
In this work we investigate the Higgs pair production in the MSSM and NMSSM
at the photon-photon collision of the ILC. We consider various experimental constraints
and scan over the parameter space of the MSSM and NMSSM. Then we calculate
the cross section of Higgs pair production  in the allowed parameter
space and compare the results with the predictions in the SM.
We find that the large enhancement of the cross
section in the MSSM is mainly due to the contributions from the loops mediated by the stau,
while in the NMSSM it is mainly due to the contributions from the top-squark loops.
For light $m_{\tilde{\tau}_1}$ and large $\mu\tan\beta$, the production rate
can be enhanced by a factor of 18 in the MSSM (relative to the SM prediction).
And for a large trilinear soft breaking parameter $A_t$ and a moderately light
top-squark, it can also be enhanced by a factor of 2 in the NMSSM.
Moreover, we also calculate the $\chi^2$ values with the LHC Higgs data and
display the results for the parameter space with $\chi^2$ better than the SM value.

\end{abstract}


\maketitle

\section{Introduction}
Since the observation of a resonance near 125 GeV at the LHC in July 2012 \cite{1207ATLAS-CMS},
both the ATLAS and CMS collaborations have accumulated more data and updated their Higgs
search results \cite{ATLAS13,CMS13}.
The combined data corresponding to the integrated luminosities
of $\sim 5 fb^{-1}$ at 7 TeV and $\sim 20 fb^{-1}$ at 8 TeV showed that
the observed particle has properties
roughly consistent with the Standard Model (SM) Higgs boson.
However, the excess in the di-photon signal rate
with respect to the SM prediction reported by the ATLAS collaboration
may indicate possible physics beyond the SM, such as supersymmetry
\cite{Feb-Cao,1213-125GeV-Higgs}.
So the next important goal of the LHC is to precisely measure its properties,
which is also the prime motivation for the future International Linear
Collider (ILC) \cite{Peskin:2012we,Ginzburg:2009aa}.
With accurate measurement of the Higgs
self-couplings, including the trilinear self-couplings and the quartic self-couplings
at the ILC, the Higgs potential can be reconstructed properly.
And the trilinear Higgs self-couplings can be measured directly in the
Higgs pair production at the ILC through the double Higgs-strahlung process
$e^+e^-\to Zhh$ and WW fusion process $e^+e^-\to \nu\nu hh$,
which have been investigated comprehensively in \cite{eemode}.
As an option of the ILC, the photon-photon collider has been paid more attention recently.
At the photon-photon collider, the Higgs pair production
is one-loop induced process with the new physics contributions and the SM contributions
at the same perturbative level, thus the cross section of the Higgs pair production may be
enhanced significantly in new physics models. Therefore, the study of Higgs pair production
at the photon-photon collider may be a sensitive probe to new physics models.

In the SM the Higgs pair production at a photon-photon collider (i.e.$\gamma\gamma\to hh$)
proceeds through one-loop triangle and box diagrams
induced by the heavy quarks and $W^{\pm}$ bosons \cite{Jikia:1992mt}.
This process may also be a sensitive probe for new physics and
has been studied in various new physics models, such as
2HDM \cite{Zhu:1997nz,Cornet:2008nq}, the vector-like quark model \cite{Asakawa:2010xj}
and the supersymmetric (SUSY) models \cite{Zhou:2003ss}.
The cross section of $\gamma\gamma\to hh$ in these new physics models
can deviate significantly from the SM prediction due to the one-loop correction
to the trilinear Higgs self-couplings \cite{Asakawa:2010xj}.
For the Higgs boson near 125 GeV, the most efficient detectable channel at the
photon-photon collider is
$\gamma\gamma\to hh\to b\bar{b}b\bar{b}$. The backgrounds, such as
$\gamma\gamma\to W^+W^-$, $ZZ$ and $b\bar{b}b\bar{b}$, can be suppressed if correct
assignment of tracks to parent partons and appropriate invariant-mass and angular
cut are achieved~\cite{Kawada:2012uy,Belusevic:2004pz}.

With the updated experimental data at the LHC, the constraints on the parameter space of
SUSY models have been becoming more and more stringent. Therefore, motivated by the latest
experimental results, we assume a SM-like Higgs boson in 123-127 GeV and
study its pair production at the photon-photon
collider in the Minimal Supersymmetric Standard Model (MSSM) \cite{MSSM}.
In the MSSM, the pair production of the SM-like Higgs boson
receives additional contributions from loops mediated by the third generation squarks,
staus, the charginos, and also the charged Higgs bosons.
It was found that the cross section of $\gamma\gamma\to hh$ can be enhanced
due to the non-decoupling effect and the mixing between the left-handed and
right-handed stau.
In the Next-to-Minimal Supersymmetric Standard Model (NMSSM) \cite{NMSSM}, which
is more favored by the experimental data \cite{July-Cao},
the studies of $\gamma\gamma\to hh$ are still absent.
Therefore, it is necessary to investigate the SM-like Higgs pair production in the NMSSM and
compare the predictions with the MSSM results.

This work is organized as follows. We first briefly describe the features of the Higgs sector
in the MSSM and NMSSM in Section II.
Then we present our numerical results for the Higgs pair production in both models in
Section III. Finally, we draw our conclusions in Section IV.

\section{Higgs sector in MSSM and NMSSM}\label{HiggsSector}
As the most economical realization of supersymmetry in particle physics, the MSSM
consists of two Higgs doublet $\hat {H_u}$ and $\hat {H_d}$. Its superpotential
has the form as follows,
\begin{eqnarray}
 W_{\rm MSSM}= W_F+\mu \hat{H_u}\cdot \hat{H_d}
 =Y_u\hat{Q}\cdot\hat{H_u}\hat{U}-Y_d \hat{Q}\cdot\hat{H_d}\hat{D}
-Y_e \hat{L}\cdot\hat{H_d} \hat{E} + \mu \hat{H_u}\cdot \hat{H_d}, \label{MSSM-pot}
\end{eqnarray}
where $\hat{Q}$, $\hat{U}$ and $\hat{D}$ denoting the squark superfields,
$\hat{L}$ and $\hat{E}$ denoting slepton superfields, and $Y_i$ ($i=u,d,e$) being
the corresponding Yukawa coupling coefficients. After the electroweak symmetry breaking,
the MSSM predicts two physical CP-even Higgs bosons $h$ and $H$, one physical
CP-odd Higgs boson $A$ and a pair of charged Higgs bosons $H^\pm$. At tree level,
this Higgs sector is determined by the mass of the CP-odd Higgs $m_A$ and the ratio
of the Higgs vacuum expectation values $\tan\beta\equiv\frac{v_u}{v_d}$.
In most cases of the MSSM, the lightest Higgs boson $h$ is SM-like (with the largest
coupling to vector bosons), and for large $m_A$ and moderate $\tan\beta$, the mass is
given by\cite{Carena-Higgsmass}
\begin{equation}\label{mh}
 m^2_{h}  \simeq m^2_Z\cos^2 2\beta +
  \frac{3m^4_t}{4\pi^2v^2} \left[\ln\frac{m^2_{\tilde t}}{m^2_t} +
\frac{X^2_t}{m^2_{\tilde t}} \left( 1 - \frac{X^2_t}{12m^2_{\tilde t}}\right)\right],
\end{equation}
with $v=174$ GeV, $m_{\tilde t} = \sqrt{m_{\tilde{t}_1}m_{\tilde{t}_2}}$
($m_{\tilde{t}_1}$ and $m_{\tilde{t}_2}$ denote the stop masses),
$X_t \equiv A_t - \mu \cot\beta$ ($A_t$ denotes the trilinear Higgs-stop coupling).
Obviously, in order to lift the Higgs boson mass up to about 125 GeV, large
$m_{\tilde t}$ or $X_t$ is needed, which in turn usually requires a large $|A_t|$.

Since the MSSM suffers from $\mu$-problem and large $m_{\tilde t}$ and $|A_t|$ induce
some extent of fine-tuning, the NMSSM has been intensively studied.
Its superpotential is given by
\begin{eqnarray}
 W_{\rm NMSSM}=W_F + \lambda\hat{H_u} \cdot \hat{H_d} \hat{S}
 + \frac{1}{3}\kappa \hat{S^3},
\end{eqnarray}
with $\hat{S}$ being singlet Higgs superfield and dimensionless parameters
$\lambda$ and $\kappa$ denoting the coupling strengths of Higgs self-interactions.
Note that when the singlet field $\hat {S}$ develops a vacuum expectation value $s$, an
effective $\mu$-term is generated by $\mu_{eff} = \lambda s$.
Compared with the MSSM, the NMSSM predicts one more CP-even Higgs boson and one
more CP-odd Higgs boson.

Due to the coupling $\lambda\hat{H_u} \cdot \hat{H_d} \hat{S}$
in the superpotential, there is additional tree level contribution to the SM-like
Higgs boson mass, i.e. $m_{h,tree}^2 = (m_Z^2 - \lambda^2 v^2 ) \cos^2 2 \beta + \lambda^2 v^2$.
Moreover, the mixing between the singlet and doublet Higgs fields can significantly alter
the Higgs boson mass. Affected by the above two factors, for $\lambda\sim 0.7$ and
$\tan\beta\sim 1$, the mass of the SM-like Higgs boson
can reach 125 GeV even without the radiative correction, which
can significantly ameliorate the fine-tuning suffered by the MSSM.

In the limit $\lambda,\kappa\rightarrow 0$ and $\mu$ is fixed, the singlet field decouples
from the doublet Higgs sector so that the NMSSM phenomenology reduces to the MSSM.
So in order to compare the Higgs sector between the two models, we require
$0.53\leq \lambda \leq 0.7$ in the NMSSM and consider two scenarios:
\begin{itemize}
\item NMSSM1 scenario: The lightest Higgs boson acts as the SM-like Higgs boson $h$. In this
scenario, the mixing effect is to pull down $m_h$, and if the mixing effect is dominant,
large radiative correction is needed to predict $m_h \simeq 125 {\rm ~GeV}$.
\item NMSSM2 scenario: The next-to-lightest Higgs boson acts as the SM-like Higgs boson $h$.
In this scenario, the mixing effect is to push up $m_h$. Both the mixing effect and the
additional tree level contribution make the large radiative correction unnecessary.
\end{itemize}

\section{Calculations and numerical results}
At the ILC, the photon-photon collider can be achieved from Compton backscattering
of laser photon off $e^+e^-$ beams, so the total cross section of
$e^+e^-\to\gamma\gamma\to hh$ can be obtained in the form
\begin{equation}\label{cross1}
  \sigma(s)=\int_{2m_h/\sqrt{s}}^{x_{max}}dz
        \frac{dL_{\gamma\gamma}}{dz}\hat{\sigma}(\gamma\gamma\to hh)\qquad \mbox{at $\hat{s}=z^2s$}
\end{equation}
where $\sqrt{s}$ ($\sqrt{\hat{s}}$) is the center of mass energies of $e^+e^-$ ($\gamma\gamma$),
and $dL_{\gamma\gamma}/dz$ is distribution function of photon luminosity, which is defined as
\begin{equation}\label{cross2}
  \frac{dL_{\gamma\gamma}}{dz}=2z\int_{z^2/x_{max}}^{x_{max}}
        \frac{dx}{x}f_{\gamma/e}(x)f_{\gamma/e}(z^2/x)
\end{equation}
For the unpolarized initial electrons and laser photon beams, the energy spectrum
of the backscattered photon is given by\cite{photonmode}
\begin{equation}\label{strfunction}
  f_{\gamma/e}(x)=\left\{
  \begin{array}{ll}
    \frac{1}{1.8397}\left(1-x+\frac{1}{1-x}-\frac{4x}{\xi(1-x)}+\frac{4x^2}{\xi^2(1-x)^2}\right) &\mbox{for $x<0.83,\quad \xi=2(1+\sqrt{2}),$}\\
    0 &\mbox{for $x>0.83$}
  \end{array}
  \right.
\end{equation}

\begin{figure}[thbp]
\includegraphics[width=13cm]{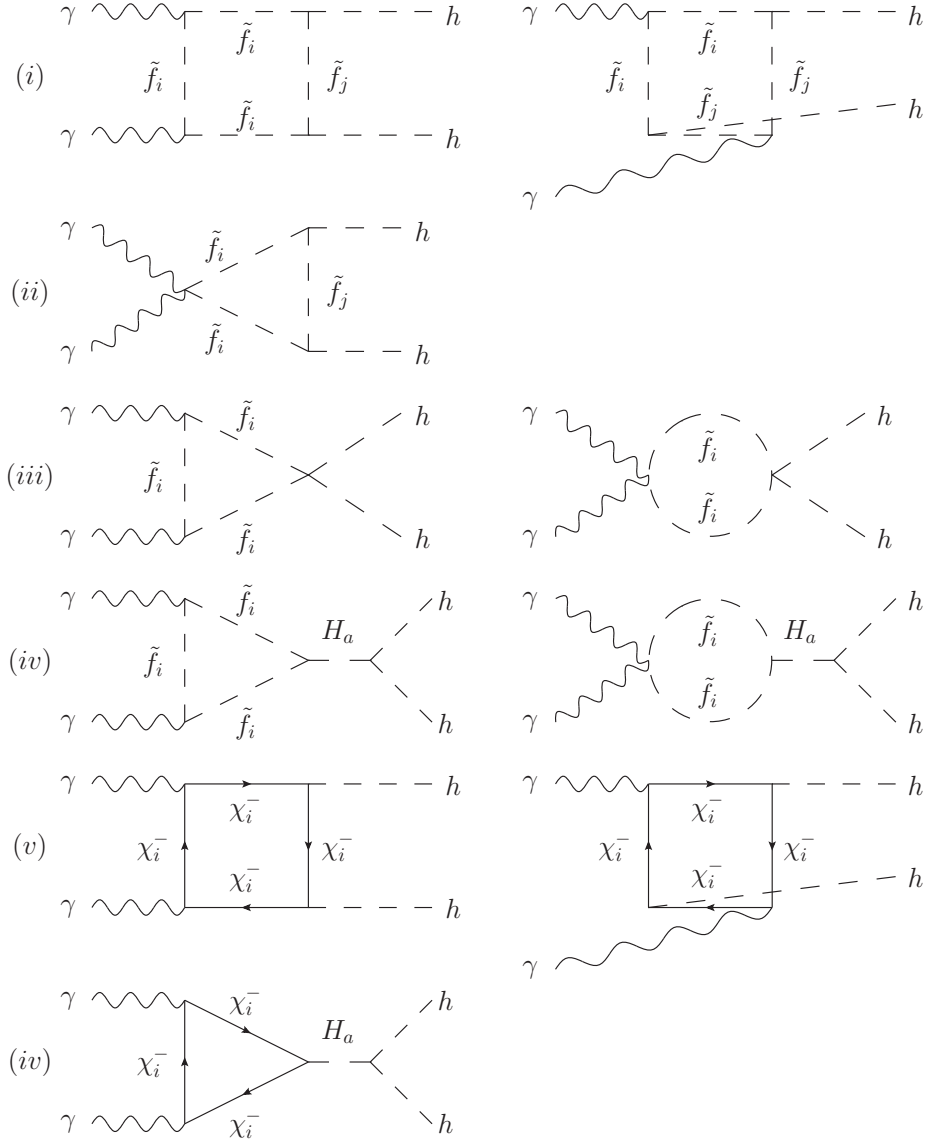}\hspace{0.5cm}
\vspace{-0.5cm}
\caption{Feynman diagrams arising from the sfermions
and charginos for the pair production of the SM-like Higgs boson at the
photon-photon collider in the MSSM and NMSSM, where $H_a$ denotes
a CP-even Higgs ($a=1,2$ for the MSSM and $a=1,2,3$ for the NMSSM),
$\tilde f_{i,j}$ ($i,j=1,2$) denotes a squark or a stau and $\chi^-_i$~($i=1,2$)
stands for a chargino. The diagrams with initial photons or final Higgs
bosons interchanged are not shown here.}
\label{fig-squark}
\end{figure}

In SUSY models the SM-like Higgs pair production at the photon-photon collider
receives additional contributions from the loops mediated by the third generation
squarks, staus, charginos and the charged Higgs boson, which are at the same perturbative
order with the SM contributions. Therefore, the cross section of the Higgs pair
production in SUSY may be enhanced significantly with respect to that in the SM.
In the calculations, we adopt the $'$t Hooft Feynman gauge~\cite{'tHooft:1971rn},
which involves a large number of diagrams from the loops mediated by W bosons,
charged Higgs bosons, the Goldstone particles and also the ghost particles,
so we do not present the Feynman diagrams from these particles, which can be
seen in \cite{Zhu:1997nz,Zhou:2003ss}, and label these diagrams as the so-called
W-C diagrams. We only show the Feynman diagrams arising
from the third generation squarks, staus and charginos in Fig.~\ref{fig-squark},
which can be divided into six parts labeled from (i) to (vi), and each
part is UV finite and gauge invariant. The calculations of the loop diagrams are
usually tedious if one expands the tensor loop functions in terms of scalar loop
functions. So we retain the tensor loop functions and use the improved
LoopTools\cite{looptools} to calculate them. In Fortran code, we use
arrays to encode the tensor loop functions as well as other quantities
such as Lorentz vectors, Dirac spinors and Dirac $\gamma$ matrices\cite{Cao:2007dk}.
The analytical expressions are so lengthy that we do not presented explicit forms here.

In the numerical calculations we take $m_t=173.5$~GeV, $m_b=4.18$~GeV,
$m_\tau$=1.78 GeV $m_W=80.385$~GeV
and $\alpha=1/128$~\cite{PDG}, and fix the center of mass energy of ILC to be 1 TeV.
For $m_h=125$~GeV, the cross section of $e^+e^-\to\gamma\gamma\to hh$ in the SM is 0.63 fb,
which changes little for $m_h$ varies from 123~GeV to 127~GeV.
And we also numerically checked our results, which are agreement with \cite{Jikia:1992mt}.

In this work, we use the package NMSSMTools \cite{NMSSMTools} to scan over the
parameter space of the MSSM and NMSSM, and investigate the samples which predict
a SM-like Higgs boson within $125\pm2$ GeV. The scan ranges of the parameter space
are same as \cite{July-Cao}, and the surviving samples satisfy the
following experimental constraints: (1)the LHC constraints on the non-standard
Higgs boson and the mass of sparticles; (2) the $2\sigma$ limits from the muon anomalous
magnetic moment, the electroweak precision data and various B-physics observables,
such as the latest experimental result of $Br(B_s\to \mu^+\mu^-)$ \cite{Bsmumu};
(3) the constraints from dark matter relic density ($2\sigma$ range given by the WMAP)
as well as the direct search result from XENON2012 experiment (at 90\% confidence level);
(4) the global fit of the SUSY predictions on various Higgs signals to the LHC Higgs
data \cite{Giardino:2012dp,Espinosa:2012ir,Ellis:2013lra,Giardino:2013bma}.
 For each surviving samples, we calculate the Higgs pair production rate
$R\equiv\sigma_{SUSY}(e^+e^-\to\gamma\gamma\to hh)/\sigma_{SM}(e^+e^-\to\gamma\gamma\to hh)$,
which is less sensitive to higher order QCD corrections.

\begin{figure}
\includegraphics[width=15cm]{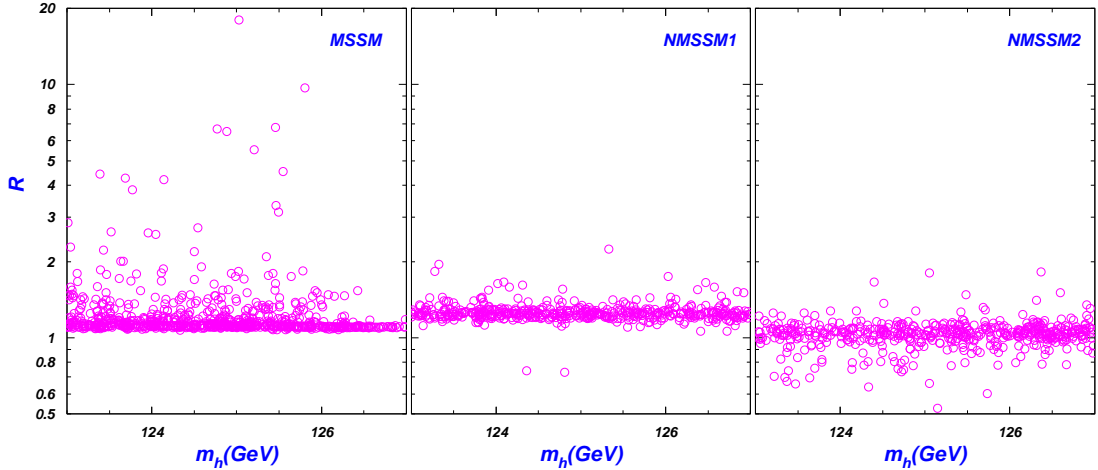}
\vspace{-0.5cm} \caption{The scatter plots of the surviving samples in the MSSM and
NMSSM, projected in the plane of $R\equiv\sigma_{SUSY}/\sigma_{SM}$ versus
the mass of SM-like Higgs boson.}
\label{fig-rtomh}
\end{figure}

In Fig.~\ref{fig-rtomh} we project the surviving samples of the MSSM and NMSSM in the plane of
$R\equiv\sigma_{SUSY}/\sigma_{SM}$ versus the mass of SM-like Higgs boson. For the case
in the NMSSM, we show the results for the NMSSM1 and NMSSM2 scenario separately.
The figure shows that in most cases the cross sections of Higgs pair production are slightly
enhanced in the MSSM and NMSSM with respect to that in the SM. However, in some special cases,
the normalized production rate can reach $\sim$18 in the MSSM and $\sim$2 in the NMSSM.
From the figure we can also see that the production rates in the NMSSM1 scenario usually
slightly larger than that in the NMSSM2 scenario, and in the NMSSM2 scenario the production
rates can also be suppressed. The reasons will be explained later.

\begin{figure}
\includegraphics[width=8cm]{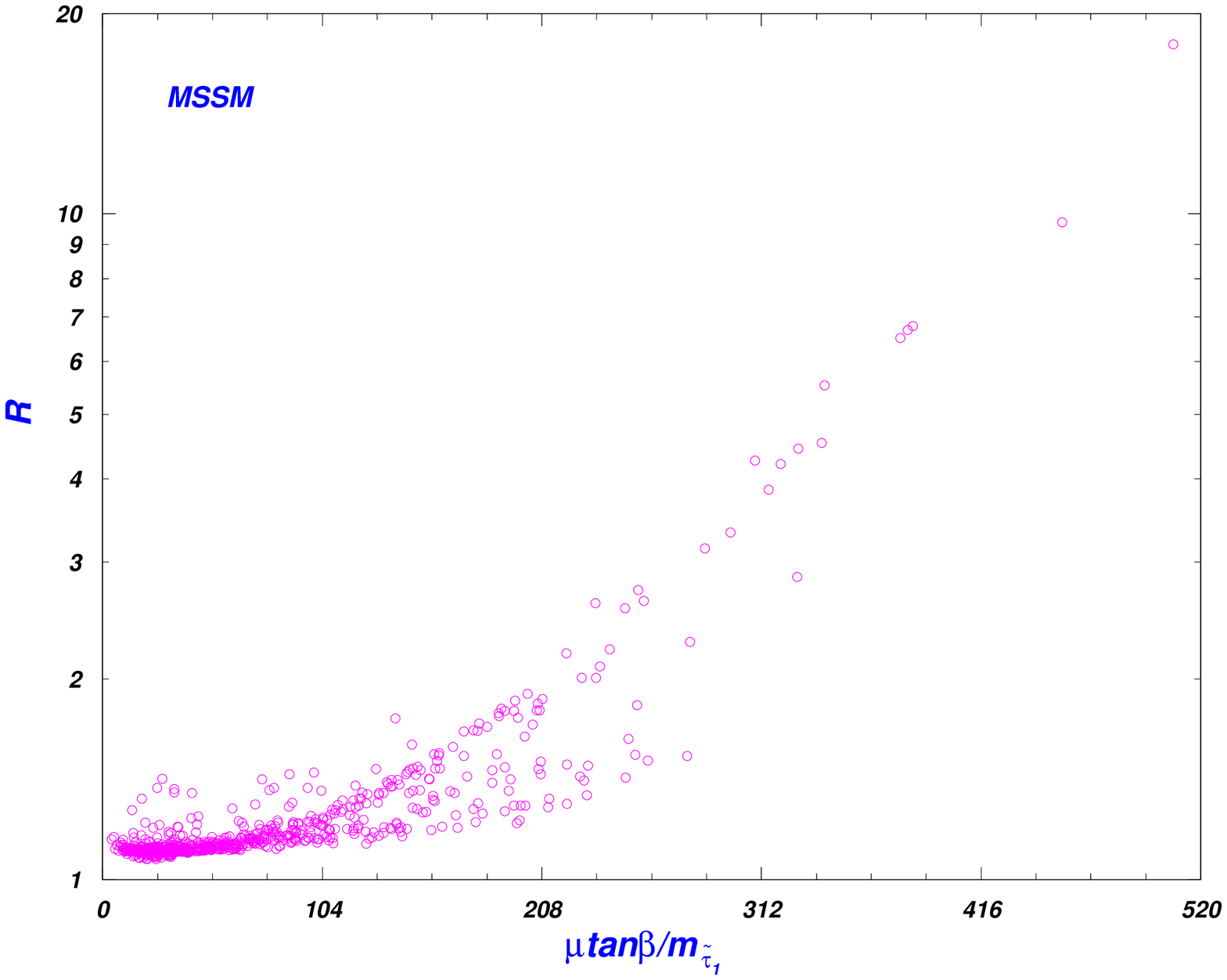}\hspace{-0.8cm}
\includegraphics[width=8cm]{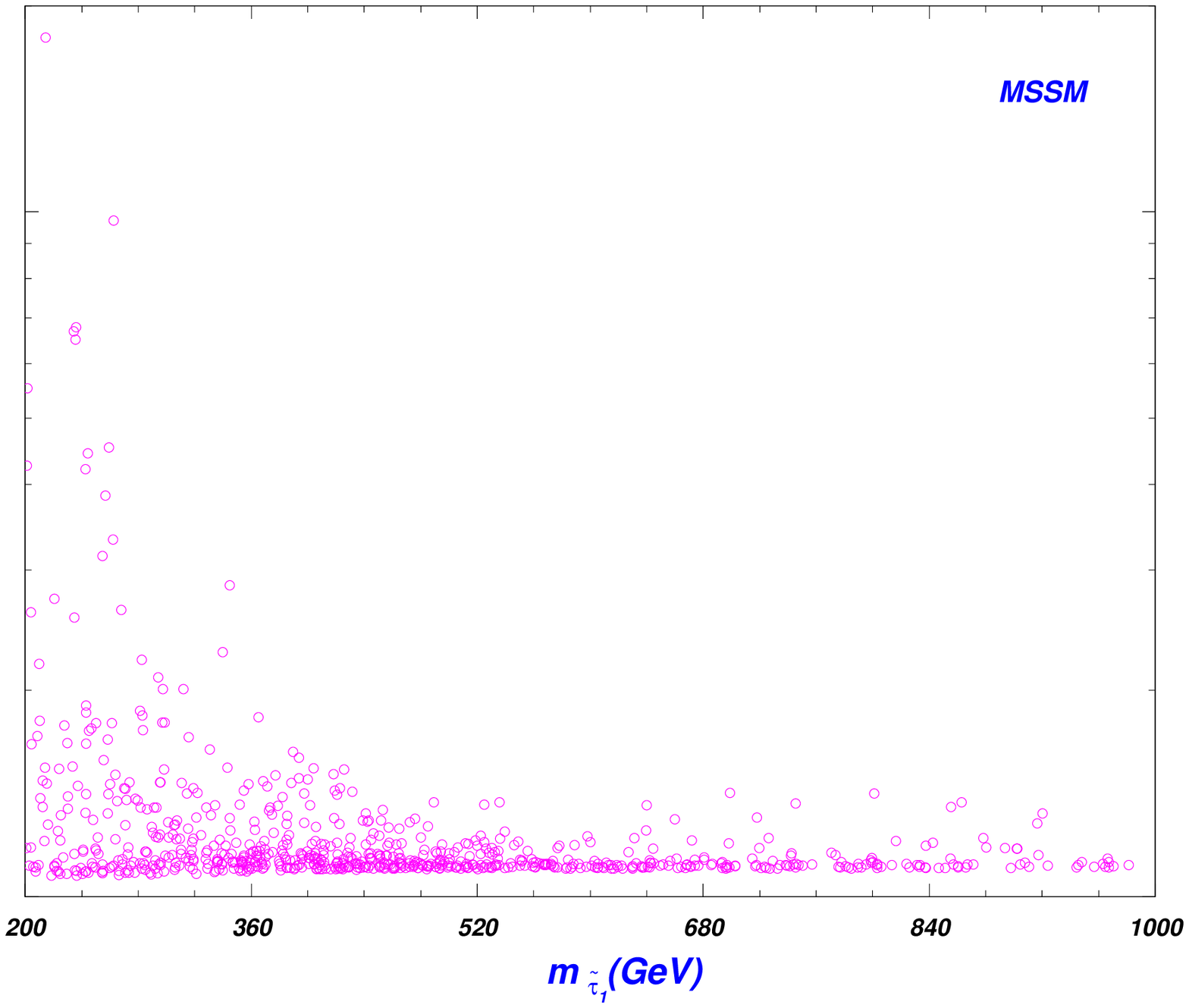}
\vspace{-0.5cm}
\caption{Same as Fig.~\ref{fig-rtomh}, but showing the surviving samples in the MSSM
in the plane of $R$ versus $\frac{\mu\tan\beta}{m_{\tilde{\tau}_1}}$ and
versus $m_{\tilde{\tau}_1}$ respectively.}
\label{fig-rto-mstau}
\end{figure}

Compared the normalized production rate of process $e^+e^-\to\gamma\gamma\to hh$
at the ILC with the process $pp\to gg\to hh$ at the LHC\cite{Cao:2013si},
we find the normalized production rate of $e^+e^-\to\gamma\gamma\to hh$
can be much larger than that of $pp\to gg\to hh$ in the MSSM, while it is usually relatively
smaller in the NMSSM. In the MSSM, it is the contributions
from the stau loops, which are absent for the process $pp\to gg\to hh$,
that may be large enough to enhance the normalized production rate of
$e^+e^-\to\gamma\gamma\to hh$ significantly.
While in the NMSSM, it is because the negligible contributions from the stau loops
and the destructive interference between the contributions from the squark loops and W boson loops
in the process $\gamma\gamma\to hh$, in which the contributions from the W boson loops may be dominant.
From the numerical calculations we also find that the contributions from the chargino loops
and bottom-squark loops to the cross section of $e^+e^-\to\gamma\gamma\to hh$ are quite small.
And the large enhancement of the cross section in the MSSM is mainly due to the contributions from
the stau loops, while in the NMSSM it is mainly due to the contributions from the top-squark loops,
which is similar with the process $pp\to gg\to hh$ at the LHC.

\begin{figure}
\includegraphics[width=15cm]{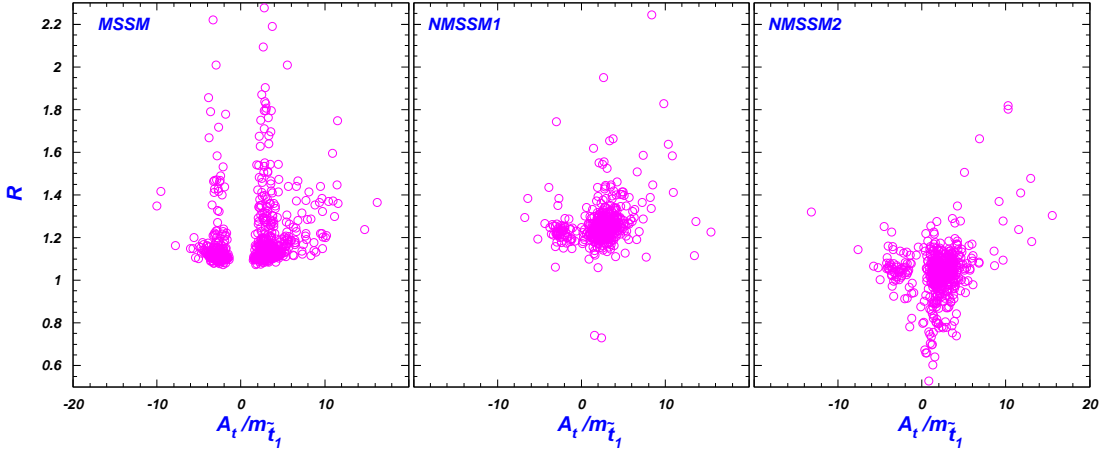}
\vspace{-0.5cm}
\caption{Same as Fig.~\ref{fig-rtomh}, but showing $R$ versus $\frac{A_t}{m_{\tilde{t}_1}}$.}
\label{fig-rto-atmsq11}
\end{figure}

\begin{figure}
\includegraphics[width=15cm]{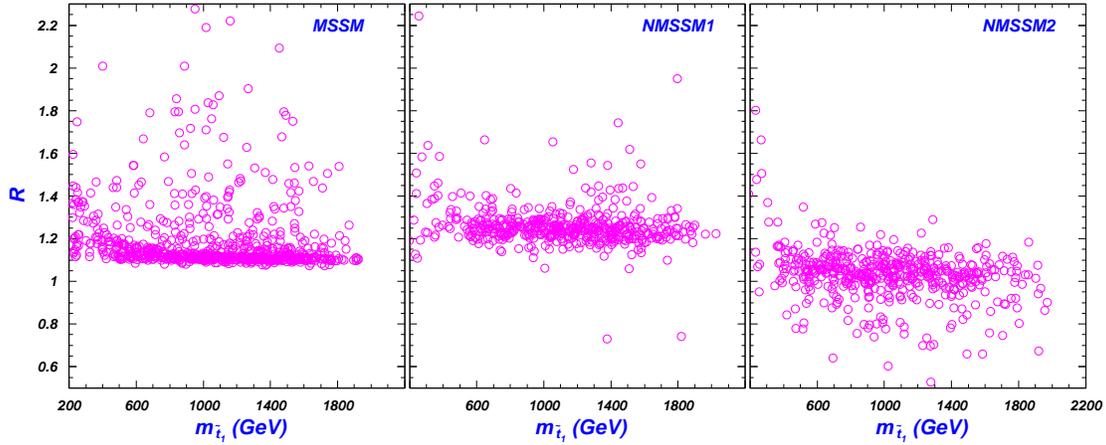}
\vspace{-0.5cm}
\caption{Same as Fig.~\ref{fig-rtomh}, but showing $R$ versus $m_{\tilde{t}_1}$.}
\label{fig-rto-msq11}
\end{figure}

For the large normalized production rate $R$ in the MSSM, its main SUSY contributions
come from the stau loops with chiral flipping (i.e. the diagrams (i) and (ii)
in Fig.\ref{fig-squark}). For light $m_{\tilde{\tau}_1}$, the amplitudes of these diagrams scale like
$(\mu\tan\beta/m_{\tilde{\tau}_1})^2$. So for light $m_{\tilde{\tau}_1}$ and large
$\mu\tan\beta$, the normalized production rate in the MSSM can be enhanced significantly,
which can be seen clearly in Fig.\ref{fig-rto-mstau}.
Compared the contributions
from the stau loops, the contributions from top-squark loops in the MSSM are relatively small.
However, it can still enhance the cross section slightly. While in the NMSSM,
due to the small $\mu\tan\beta$, the contributions from the stau loops are negligible
and the effect of top-squark loops is remarkable. Therefore,
in Fig.~\ref{fig-rto-atmsq11} and Fig.\ref{fig-rto-msq11}
we show the normalized production rate $R$ as a function of $\frac{A_t}{m_{\tilde{t}_1}}$
and $m_{\tilde{t}_1}$, respectively.

From the Fig.\ref{fig-rto-atmsq11} and Fig.\ref{fig-rto-msq11} we can see that
large $\frac{|A_t|}{m_{\tilde{t}_1}}$ and light $m_{\tilde{t}_1}$ usually
predict a large normalized production rate $R$ in the NMSSM.
This is because for a light $m_{\tilde{t}_1}$, the dominant contributions from
top-squark loops scale like
$(A_t/m_{\tilde{t}_1})^4(m_t^2m_{\tilde{t}_1}^2/m_{\tilde{t}_2}^4)$\cite{Cao:2013si}.
Fig.\ref{fig-rto-msq11} also indicates that in a corner of the parameter space in the NMSSM1
scenario, the deviation can be larger than 40\% for $m_{\tilde{t}_1}\sim$ 1.5 TeV.
In such cases, the contributions from the so-called W-C diagrams
 are usually larger than the cases in the NMSSM2 scenario.
In fact, it is the contributions from the charged Higgs boson loops and Goldstone boson
loops that lead to the larger enhancement in the NMSSM1 scenario. Different from
the couplings in the SM, the coupling of SM-like Higgs boson with charged Higgs
boson $H^\pm$ or Goldstone boson $G^\pm$ in the NMSSM has additional interactions from
 the Higgs singlet field \cite{Franke:1995tc},
\begin{eqnarray}
C_{hH^+H^-}&\sim& U_{a3}^S[(2\kappa\mu_{eff}+\lambda A_\lambda)\sin2\beta+2\lambda\mu_{eff}] \nonumber\\
C_{hG^+G^-}&\sim& U_{a3}^S[-(2\kappa\mu_{eff}+\lambda A_\lambda)\sin2\beta+2\lambda\mu_{eff}]
\label{req02}
\end{eqnarray}
where $U_{a3}^S$ denotes the singlet component of the SM-like Higgs boson.
Note that NMSSM1 scenario usually prefers larger $\mu_{eff}$ and smaller $\tan\beta$ than
NMSSM2 scenario\cite{Feb-Cao}. And for these samples, we have numerically checked that the couplings
$C_{hH^+H^-}$ and $C_{hG^+G^-}$ are usually larger in the NMSSM1 scenario than that in the NMSSM2 scenario.

Note that even for heavy $m_{\tilde{t}_1}$, the normalized production rate in the MSSM
can also be enhanced slightly. This is mainly due to the contributions from top-squark loops.
For heavy $m_{\tilde{t}_1}$, the amplitude of the dominant diagrams from top-squark
loops can be written as
\begin{equation}
M \sim \alpha_s^2 Y_t^2 ( c_1\sin^22\theta_t\frac{A_t^2}{m_{\tilde{t}_1}^2}+c_2\frac{A_t^2}{m_{\tilde{t}_2}^2} )\label{req01}
\end{equation}
with $Y_t$ denotes the top quark Yukawa coupling, $\theta_t$ is the chiral mixing angle
and $c_1$ and $c_2$ are $\cal{O}$(1) coefficient with opposite signs.
In this case, the mass splitting between $m_{\tilde{t}_1}$ and $m_{\tilde{t}_2}$ is
small (i.e. $\theta_t\sim\frac{\pi}{4}$), so the two terms in Eq.(\ref{req01}) cancel severely.
However, because $|A_t|$ is usually larger than stop mass, the cross section of Higgs pair
production in the MSSM can still be enhanced by about 10\%.

As analyzed in the section II, in order to predict a 125 GeV Higgs in the MSSM,
the average stop mass $m_{\tilde{t}}$ or the trilinear soft breaking parameter $|A_t|$
must be large. However, in the NMSSM1 scenario, the mixing effect is destructive with
the additional tree level contribution, so large radiative corrections are needed
to predict a 125 GeV Higgs boson. For the same values of the top-squark in the
NMSSM, the NMSSM1 scenario usually prefers a larger $|A_t|$, which leads to the larger
production rate than the NMSSM2 scenario, as shown in Fig.~\ref{fig-rtomh}.
In the NMSSM2 scenario, both the additional
tree level contribution and the mixing effect can enhance the Higgs boson mass up to
about 125 GeV, so the constraints on the parameter $|A_t|$ is not so strong. That is why
the production rate can also be enhanced or suppressed in the NMSSM2 scenario.

\begin{figure}
\includegraphics[width=15cm]{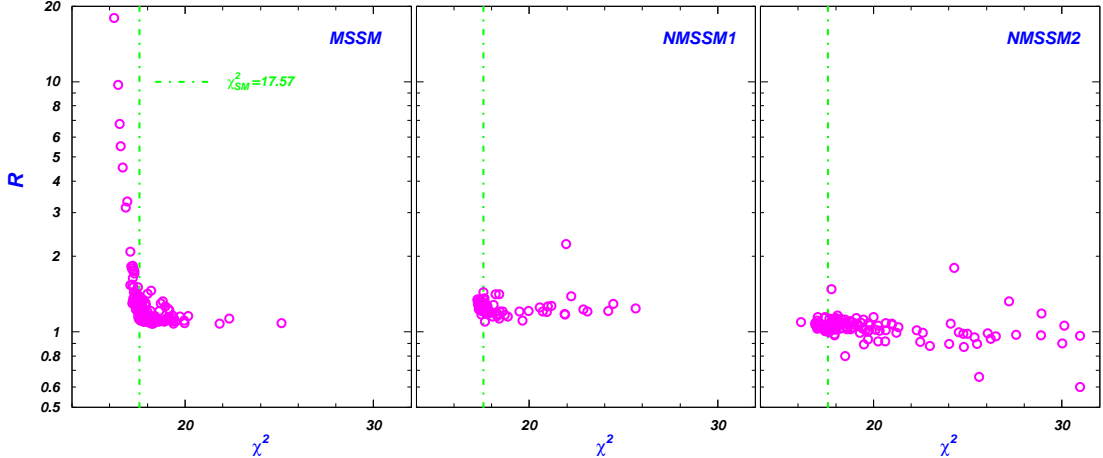}
\vspace{-0.5cm}
\caption{Same as Fig.~\ref{fig-rtomh}, but showing $R$ versus $\chi^2$.
Here only the samples satisfying 125~GeV$\le m_h\le$ 126~GeV are plotted.}
\label{fig-rtochi2}
\end{figure}

Finally, we calculate the $\chi^2$ values with the LHC Higgs data for the samples
with 125~GeV$\le m_h\le$126~GeV in the best fitted mass
region~\cite{Giardino:2012dp,Espinosa:2012ir,Ellis:2013lra,Giardino:2013bma}, and show the
relationship with the normalized production rate $R$ in Fig.~\ref{fig-rtochi2}.
From the figure we can see that there exist some samples with $\chi^2$ slightly
smaller than its SM value ($\chi^2_{SM}=17.57$).
For the $\chi^2$ better than the SM value, the production rates only deviate slightly
from the SM prediction in the NMSSM2 scenario, while the deviation can reach
40$\%$ in the NMSSM1 scenario and the production rate in the MSSM can be enhanced about
18 times larger than the SM prediction.
And we also numerically check that, for the samples with $\chi^2$ much larger than its SM value,
the coupling of Higgs boson to b quark usually deviates significantly from its SM prediction,
or/and the singlet component of the SM-like Higgs boson is usually large.

\section{Summary and Conclusion}\label{Sum}
For the precise measurement
of the Higgs boson properties at the ILC, the Higgs pair
production at the photon-photon collision will play
an important role since it can test the Higgs self-coupling.
In this work we investigated the Higgs pair
production in the MSSM and NMSSM at the photon-photon collision of the ILC.
We considered various experimental constraints
and scanned over the parameter space of the MSSM and NMSSM. Then in the allowed parameter
space we calculated the cross section of Higgs pair production at the ILC and compared
the results with the predictions in the SM. We found that the large enhancement of the cross
section in the MSSM is mainly due to the contributions from the loops mediated by the stau,
while in the NMSSM it is mainly due to the contributions from the top-squark loops.
For light $m_{\tilde{\tau}_1}$ and large $\mu\tan\beta$, the normalized production
rate $R\equiv\sigma_{SUSY}/\sigma_{SM}$ in the MSSM can reach 18.
And for a large trilinear soft breaking parameter $A_t$ and a moderate
top-squark mass $m_{\tilde{t}_1}$, the normalized production rate $R$
 can also reach $\sim$2 in the NMSSM. We also calculated the $\chi^2$ values with the LHC Higgs data.
For the $\chi^2$ better than the SM value, the production rates only deviate slightly
from the SM prediction in the NMSSM2 scenario, while the deviation can reach
40$\%$ in the NMSSM1 scenario and the production rates in the MSSM can be enhanced about
18 times larger than the SM prediction.

\section*{Acknowledgement}
We thank Jin Min Yang, Junjie Cao and Jingya Zhu for helpful discussions.
This work was supported in part by the National Natural Science
Foundation of China (NNSFC) under grant No. 11247268.

%%%%%%%%%%%%%%%%%%%%%%%%%%%%%%%%%%%%%%%%%%%%%%%%%%%%%%%%%%%%%%%%%%%%%%%%%%%%%%

\end{document}